\documentclass[%
superscriptaddress,
preprint,
 amsmath,amssymb,
 aps,
prl,
showkeys,
]{revtex4-2}

\usepackage{graphicx}
\usepackage{dcolumn}
\usepackage{bm}
\usepackage{newunicodechar}
\newunicodechar{√}{\ensuremath{\sqrt{}}}
\usepackage{hyperref}

\begin{document}


\title{Charge order-driven nematicity in the nickel-pnictide superconductor {Ba$_{1-x}$Sr$_x$Ni$_2$As$_2$}}


\author{Thomas Johnson}
    \affiliation{Department of Physics, The Grainger College of Engineering, University of Illinois Urbana–Champaign, Urbana, 61801, IL, USA}
    \affiliation{Materials Research Laboratory, The Grainger College of Engineering, University of Illinois Urbana–Champaign, Urbana, 61801, IL, USA}
\author{Camille Bernal-Choban}
    \email{cmbc@illinois.edu}
    \affiliation{Department of Physics, The Grainger College of Engineering, University of Illinois Urbana–Champaign, Urbana, 61801, IL, USA}
    \affiliation{Materials Research Laboratory, The Grainger College of Engineering, University of Illinois Urbana–Champaign, Urbana, 61801, IL, USA}
\author{Sangjun Lee}
    \affiliation{Samsung Institute of Advanced Technology, KOR}
\author{Xuefei Guo}
    \affiliation{Department of Physics, The Grainger College of Engineering, University of Illinois Urbana–Champaign, Urbana, 61801, IL, USA}
    \affiliation{Materials Research Laboratory, The Grainger College of Engineering, University of Illinois Urbana–Champaign, Urbana, 61801, IL, USA}
\author{Stella Sun}
    \affiliation{Department of Physics, The Grainger College of Engineering, University of Illinois Urbana–Champaign, Urbana, 61801, IL, USA}
    \affiliation{Materials Research Laboratory, The Grainger College of Engineering, University of Illinois Urbana–Champaign, Urbana, 61801, IL, USA}
\author{John Collini}
    \affiliation{Maryland Quantum Materials Center, Department of Physics, University of Maryland, College Park, 20742, MD, USA}
\author{Christopher Eckberg}
    \affiliation{Maryland Quantum Materials Center, Department of Physics, University of Maryland, College Park, 20742, MD, USA}
\author{Johnpierre Paglione}
    \affiliation{Maryland Quantum Materials Center, Department of Physics, University of Maryland, College Park, 20742, MD, USA}
    \affiliation{Canadian Institute for Advanced Research, Toronto, Ontario M5G 1Z8, Canada}
\author{Rafael M. Fernandes}
    \affiliation{Department of Physics, The Grainger College of Engineering, University of Illinois Urbana–Champaign, Urbana, 61801, IL, USA}
    \affiliation{Anthony J. Leggett Institute for Condensed Matter Theory, The Grainger College of Engineering, University of Illinois Urbana-Champaign, Urbana, 61801, IL, USA}
\author{Eduardo Fradkin}
    \affiliation{Department of Physics, The Grainger College of Engineering, University of Illinois Urbana–Champaign, Urbana, 61801, IL, USA}
    \affiliation{Anthony J. Leggett Institute for Condensed Matter Theory, The Grainger College of Engineering, University of Illinois Urbana-Champaign, Urbana, 61801, IL, USA}
\author{Peter Abbamonte}
    \email{abbamonte@illinois.edu}
    \affiliation{Department of Physics, The Grainger College of Engineering, University of Illinois Urbana–Champaign, Urbana, 61801, IL, USA}
    \affiliation{Materials Research Laboratory, The Grainger College of Engineering, University of Illinois Urbana–Champaign, Urbana, 61801, IL, USA}

\date{\today}

\begin{abstract}
Nematic order refers to the spontaneous breaking of rotational symmetry while preserving translational symmetry. First identified in classical liquid crystals, nematic order arises from the collective alignment of anisotropic molecules. Its quantum counterpart, electronic nematicity, has been observed in a variety of quantum materials, ranging from unconventional superconductors to kagome metals. Despite its prevalence, there is no universal understanding of the conditions under which nematic order occurs. Electronic nematicity is most firmly established in iron-based superconductors, where it is understood to be a consequence of vestigial spin density wave (SDW) order. However, direct evidence for nematicity arising from other types of order are lacking. Here, we report direct evidence for charge-order-driven electronic nematicity in Ba$_{1-x}$Sr$_x$Ni$_2$As$_2$, a nickel-based analog of the iron pnictides known to exhibit charge density wave (CDW) order. Using x-ray diffraction under applied uniaxial strain, we observe a pronounced symmetry-breaking response---up to $\sim 50 \%$---in the intensity of incommensurate CDW Bragg peaks, even at small strain levels ($\epsilon_{xy} \sim 10^{-3}$). This effect occurs within the same region of the phase diagram where a giant nematic susceptibility is observed in transport measurements. These results provide direct evidence that long-range CDW order can drive nematic behavior in quantum materials.
\end{abstract}

\maketitle


Rotational symmetry breaking in liquid crystals influences viscoelastic, vibrational, and flow properties of these complex fluids \cite{miesowiczm.1936, frankLiquidCrystalsTheory1958, oseenc.w.1933, zocherh.1933, stephenPhysicsLiquidCrystals1974}. Such emergent behaviors, which typify classical nematicity, are driven by the spontaneous orientation of anisotropic molecules. The discovery of electronic nematicity, and its inherent intertwining of (multiple) competing electronic phases, creates analogous signatures in electrical transport and has emerged as a fundamental phenomenon in correlated electronic materials.

Since its first experimental observation in GaAs two-dimensional electron gases (2DEGs) \cite{pan_strongly_1999}, reports of electronic nematicity have expanded to include other quantum Hall systems \cite{Lilly-1999, pan_strongly_1999, Fradkin-2010, Samkharadze-2016, Qian-2017}, copper-oxides \cite{Hinkov-2008,Fradkin-2015,nakata_nematicity_2021, auvray_nematic_2019}, iron-based superconductors\,\cite{Chu2012,Yi2011,kuo_ubiquitous_2016,Worasaran2021}, correlated oxides\,\cite{borzi_formation_2007}, twisted bilayer graphene\,\cite{Cao2021}, and cold atoms in optical lattices\,\cite{jin_evidence_2021}. 

Identifying the underlying causes of electronic nematic order is challenging. The point-particle nature of electrons means that, unlike its classical counterpart, electronic nematicity cannot be attributed to a simple shape anisotropy of its constituents. 
However, in all known cases, nematic fluctuations emerge near electronic phase transitions that break one or more spatial or rotational symmetries. 

This phenomenon is most clearly established in iron pnictides \,\cite{Fernandes2022} where the nearby spin density wave order drives the nematic instability\,\cite{Fang2008,xu_ising_2008,fernandes_what_2014}. These fluctuations are manifestations of `vestigial order,' in which a secondary order parameter condenses prior to the onset of the primary one\,\cite{nie_quenched_2014,Fradkin-2015,fernandes_intertwined_2019}, producing measurable effects in the elastic shear modulus and spin-lattice relaxation rate\,\cite{fernandes_scaling_2013,Bohmer2014,lu_nematic_2014}. However, many aspects of this mechanism remain unresolved\,\cite{Bohmer2022}, and nematic order in other material systems appears to arise from entirely different origins\,\cite{ayres_transport_2022}.

It has long been proposed that charge order could drive electronic nematicity in non-magnetic materials, since charge density waves (CDWs) often break rotational symmetry\, \cite{fradkin_colloquium_2015,fernandes_intertwined_2019,oganesyan_quantum_2001,nie_vestigial_2017,agterberg_physics_2020}. In this scenario, nematic order could emerge from a $d$-wave Pomeranchuk instability of a Fermi liquid\,\cite{oganesyan_quantum_2001}, with the electronic phase transition becoming continuously intertwined with a CDW transition\,\cite{nie_quenched_2014,nie_vestigial_2017,Fradkin-2015}. Such a mechanism may arise near a CDW instability that breaks rotational symmetry by selecting one of two wave vectors related by a 90$^\circ$ rotation\,\cite{nie_vestigial_2017}.

Despite theoretical support, direct experimental evidence for an electronic nematic order parameter associated with a CDW is lacking. Some studies suggest that a charge-driven mechanism may contribute to nematicity in the cuprates, but conclusive evidence has yet to emerge\,\cite{murayamaDiagonalNematicityPseudogap2019, nie_vestigial_2017}. More recently, kagome metals have been explored as potential hosts for CDW-driven nematicity\,\cite{grandi_theory_2023,Picot2015,Lugon2019}, but experimental results are mixed\,\cite{nie_charge-density-wave-driven_2022,Jiang2024,Wu2023,asaba_evidence_2024,Liu-2024b}.

The Ba$_{1-x}$Sr$_x$Ni$_2$As$_2$ (BSNA) family of superconductors is another promising candidate for charge order-driven nematicity\,\cite{eckberg_sixfold_2020, Yao2022, Souliou2022, Song2023, Frachet2022, chen_charge_2024}. These materials are non-magnetic analogs of the 122 iron-pnictide superconductors\,\cite{eckberg_sixfold_2020, narayan_potential_2023}, exhibit electronic nematicity\,\cite{eckberg_sixfold_2020, Yao2022}, and host multiple CDWs\,\cite{lee_multiple_2021}. 
Notably, the onset of nematic order in BSNA coincides with a sixfold enhancement of the superconducting transition temperature, suggesting an intertwining between nematicity and superconductivity\,\cite{eckberg_sixfold_2020,lederer_tests_2020}.

Here, we present direct evidence for charge order-driven electronic nematicity in BSNA. Using x-ray diffraction on Ba$_{1-x}$Sr$_x$Ni$_2$As$_2$ with $x = 0.27$, we observe a pronounced anisotropic response of symmetry-related incommensurate CDW superlattice peaks under small uniaxial strain. From this anisotropy, we define a nematic order parameter and extract the charge-nematic susceptibility, which closely follows the Curie-Weiss-like divergence previously observed in elastoresistivity\,\cite{eckberg_sixfold_2020}. By fitting the CDW anisotropy to a Ginzburg-Landau model, we show a strong anharmonic coupling between charge-nematic fluctuations and strain. These results provide the first direct demonstration linking charge order to a nematic state, with important implications for the intertwined superconductivity in this material system.

All x-ray diffraction experiments were performed on BSNA $x = 0.27$, crystals grown by the self-flux method (Ref.\,\cite{sefat_structure_2009}) and characterized in Ref.\,\cite{lee_multiple_2021}. A piezoelectric strain device, identical to those employed in previous transport studies\,\cite{kuo_ubiquitous_2016,eckberg_sixfold_2020}, was used to apply uniaxial strain, $\epsilon$, along the sample's tetragonal \textbf{b}-axis, breaking C$_4$ rotational symmetry in the B$_{1g}$ channel. At each applied strain, we collected cooling and heating measurements using a Xenocs GeniX$^{3\mathrm{D}}$ Mo K$\alpha$ ($\lambda=0.7093$\AA) micro-spot x-ray source with a Mar345 image plate detector, with the sample and strain cell residing in a liquid helium cryostat with beryllium windows. Our strain value, $\epsilon$, ranged from $\sim 0$ to $1.3 \times 10^{-3}$ with an estimated error of $7 \times 10^{-5}$. Additional details of sample preparation and experimental setup are in the Supplemental Material\,\cite{supp}.

At a substitution of $x=0.27$, BSNA hosts both a commensurate (C) and an incommensurate (I) charge density wave (CDW)\,\cite{eckberg_sixfold_2020, lee_multiple_2021}. The I-CDW exists in the tetragonal phase and exhibits fourfold rotational symmetry, with wave vectors $(0.28, 0, 0)_\mathrm{tet}$ and $(0, 0.28, 0)_\mathrm{tet}$\,\cite{lee_multiple_2021}. Notably, the I-CDW appears in the same region of the phase diagram where the nematic response observed in transport is most pronounced\,\cite{eckberg_sixfold_2020}. In contrast, the C-CDW, with wave vector $(1/3, 0, 0)_\mathrm{tri}$, forms in the triclinic (low-temperature) phase, where rotational symmetry is explicitly broken, leading to a complex pattern of twin domains. Due to its relative simplicity and its connection to the nematic response observed in transport\,\cite{eckberg_sixfold_2020}, this study focuses on the I-CDW. This coherent “2Q” state retains rotational symmetry, which can be broken by applying a symmetry-breaking strain field.

We measure the temperature dependence of the I-CDW peaks along the $x$ and $y$ reciprocal space directions under various applied strain values. Specifically, we use the CDW superlattice reflections $\mathbf{Q}_x = (-0.28, 3, -5)_{\mathrm{tet}}$ and $\mathbf{Q}_y=(-1, -1.72, -5)_{\mathrm{tet}}$, which correspond to the $x$ and $y$ components of the CDW in the (0,3,-5) and (-1,-2,-5) Brillouin zones, respectively. These reflections were chosen for convenience, and since their wave vectors are related by a $90^\circ$ rotation, their relative intensity provides a direct measure of the CDW anisotropy in the $(x,y)$ plane.

\begin{figure}[ht]
	\centering
	\includegraphics[width=0.9\linewidth]{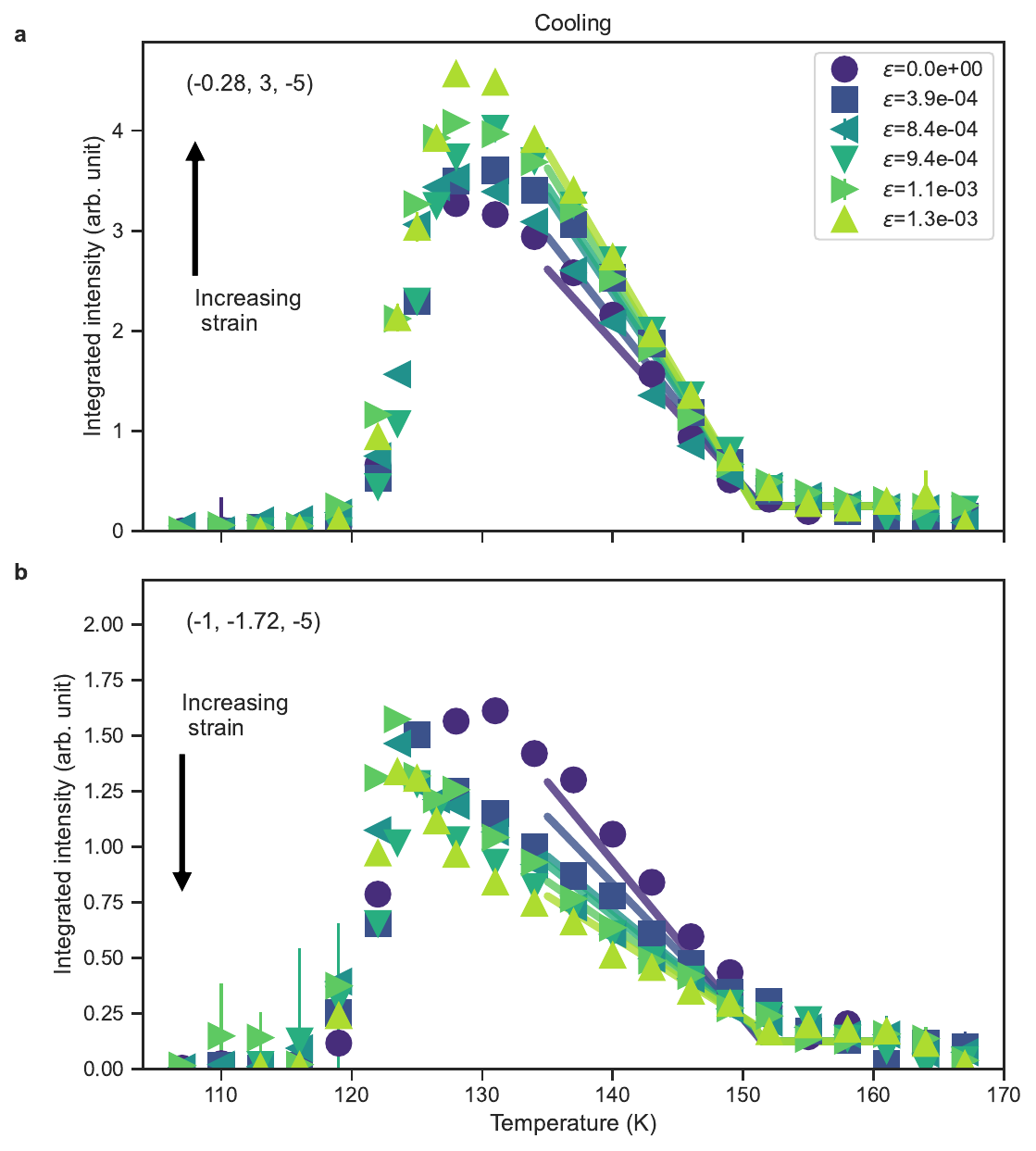}
	\caption{{\bf Temperature dependence of two I-CDW satellites, related to one another by a 90$^\circ$ rotation, for various applied strain values.} {\bf a.} Integrated intensity of the $\mathbf{Q}_x = (-0.28,\,3,\,-5)_\mathrm{tet}$ reflection, normalized to its value at $T=135$ K at zero strain. {\bf b.} Integrated intensity of the $\mathbf{Q}_y = (-1,\,-1.72,\,-5)_\mathrm{tet}$, also normalized to its value at $T=135$ K at zero strain. Solid lines are fits to the data using the Landau model described in the text. Bars correspond to standard errors from least squares fits.}
	\label{fig:icdw_t}
\end{figure}

Our main results, presented in Fig.\,\ref{fig:icdw_t}, show the integrated intensities of these two CDW reflections as a function of temperature over the range 107\,K$\,<\,T\,<\,167$\,K for several applied strain values. To account for differences in the CDW structure factors in different Brillouin zones, we normalize the intensities of both reflections to their values at $T = 135$\,K and $\epsilon$ = 0 (see Supplemental Material\,\cite{supp} Fig. S7). No hysteresis was observed between heating and cooling cycles, so only cooling data are shown. (Warming data are included in the Supplemental Material\,\cite{supp}.)

The I-CDW reflections emerge below $T_\mathrm{IC}$ = 151\,K, grow continuously with decreasing temperature, and abruptly disappear below the triclinic transition at $T_\mathrm{tri}\,\sim\,125$\,K\,\cite{lee_multiple_2021} (see Supplemental Material\,\cite{supp} Fig. S2 for details on transition temperature determination). Applying uniaxial strain increases the integrated intensity of the $\mathbf{Q}_x$ reflection while decreasing that of $\mathbf{Q}_y$, demonstrating that even small values of applied strain break the fourfold rotational symmetry of the CDW. 

\begin{figure}[ht]
	\centering
	\includegraphics[width=0.9\linewidth]{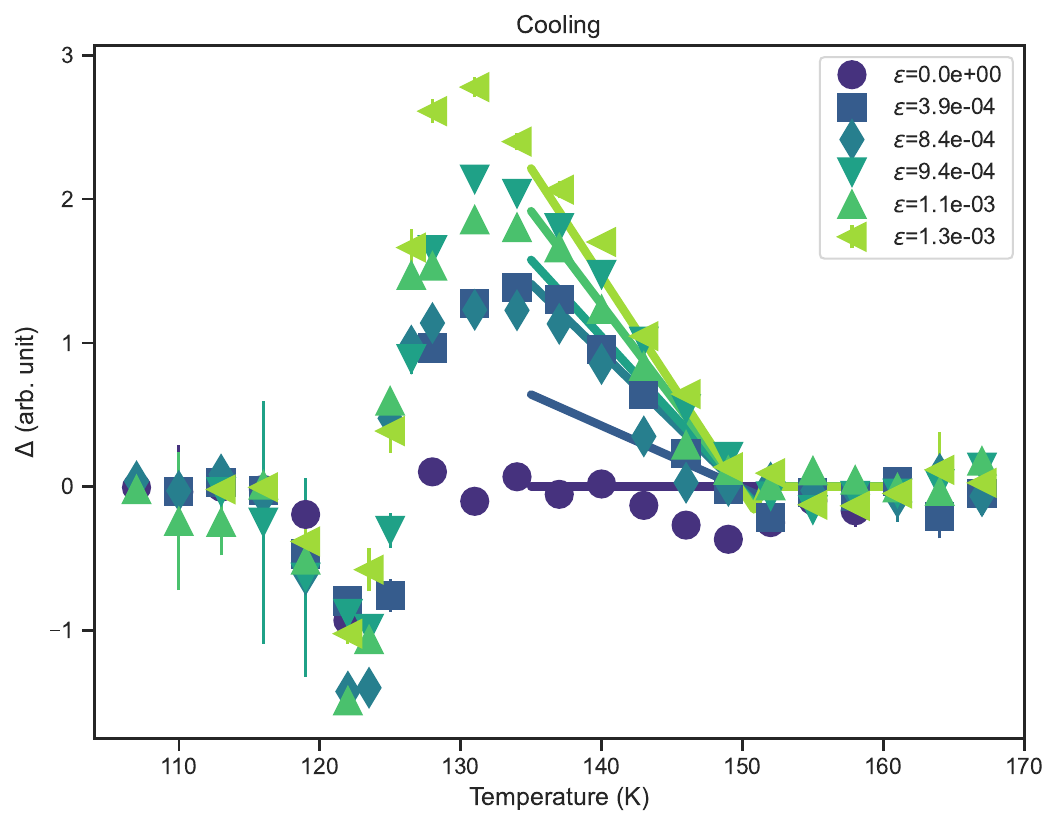}
	\caption{{\bf Temperature dependence of the nematic order parameter, $\Delta = I_x - I_y$, which represents the rotational anisotropy of the I-CDW, for various applied strain values.} The 90$^\circ$ rotational symmetry of the I-CDW is broken by even trace amounts ($\epsilon \sim 10^{-4}$) of strain. Solid lines correspond to the Landau fits described in the text. Bars represent standard errors from least squares fits.}
	\label{fig:delta}
\end{figure}

Following Ref.\,\cite{nie_quenched_2014}, we define a nematic order parameter based on the anisotropy of the I-CDW,
\begin{equation}
\Delta(\epsilon, T) = I_x(\epsilon, T) - I_y(\epsilon, T),
\end{equation}
where $I_x$ and $I_y$ denote the integrated intensities of the $\mathbf{Q}_x$ and $\mathbf{Q}_y$  reflections, respectively, normalized to their values at $T=135$\,K. This anisotropy parameter, $\Delta(\epsilon, T)$, is odd under fourfold rotations around the $c$-axis and vanishes by definition when $\epsilon = 0$. It characterizes how the rotational symmetry of the I-CDW evolves with temperature and applied strain, analogous to the resistive anisotropy investigated in Ref.\,\cite{eckberg_sixfold_2020}.
The full temperature and strain dependence of $\Delta$ is shown in Fig.\,\ref{fig:delta}. In the absence of strain, $\Delta$ remains zero except within a narrow temperature range below $T_\mathrm{tri}$, where rotational symmetry is explicitly broken by the triclinic transition. However, upon applying strain of order $\epsilon\sim10^{-4}$, we observe a nonzero $\Delta$ that emerges at $T_\mathrm{IC}$ and increases linearly with strain. This behavior becomes nonlinear below $T_\mathrm{tri}$, where the I-CDW vanishes. We do not observe a significant shift in the transition temperature $T_\mathrm{IC}$ across different strain values.

\begin{figure}[ht]
	\centering
	\includegraphics[width=0.9\linewidth]{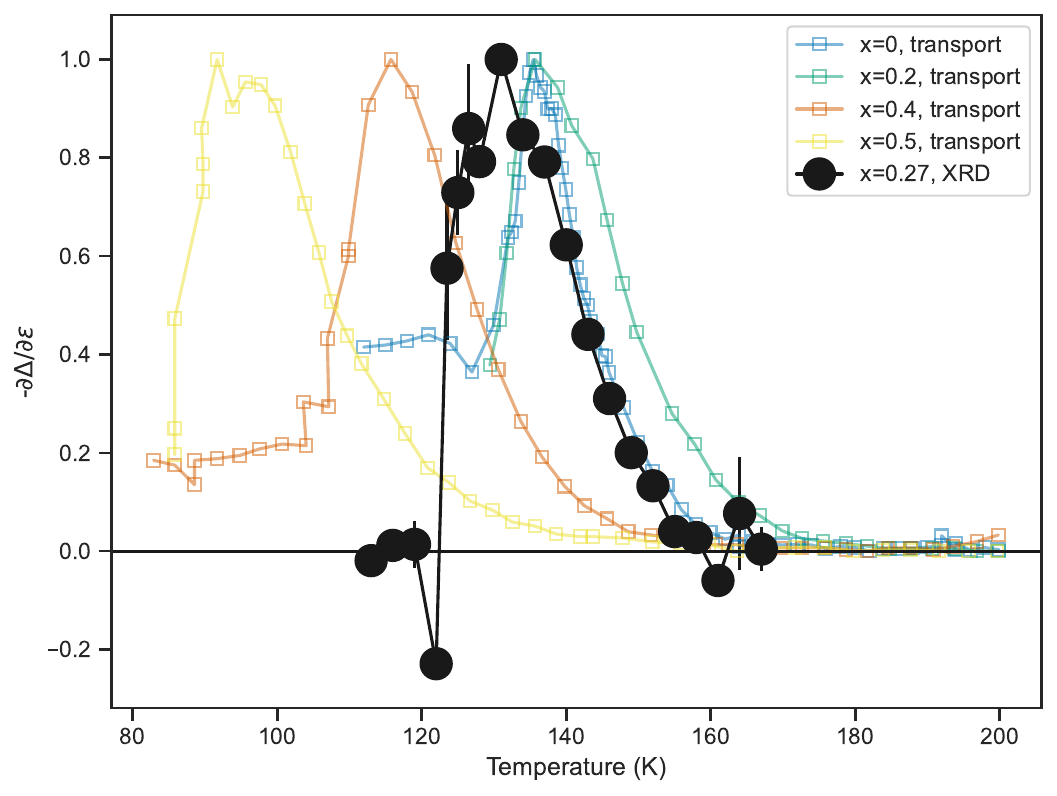}
	\caption{{\bf Nematic susceptibility, $\chi(T)$ for BSNA with $x=0.27$ (black circles) from our x-ray measurements.} The data are plotted alongside nematic susceptibility data from the elasto-resistivity measurements of Ref.\,\cite{eckberg_sixfold_2020} at similar substitutions (open squares). Remarkable consistency is seen betwen the two different techniques, suggesting the two effects have the same origin. Error bars for $x=0.27$ correspond to standard errors from least squares fits.}
	\label{fig:chi}
\end{figure}

As shown in Fig.\,\ref{fig:delta}, even small amounts of applied strain are sufficient to break the rotational symmetry of the I-CDW. Drawing an analogy to elastoresistivity studies that report a strong nematic response in BSNA \cite{eckberg_sixfold_2020}, we define a CDW-nematic susceptibility as
\begin{equation}
\chi(T) = \frac{\partial\Delta(\epsilon, T)}{\partial\epsilon},
\end{equation}
which quantifies the sensitivity of the I-CDW to uniaxial strain. The temperature dependence of $\chi(T)$, obtained from a linear fit to the data in Fig.\,\ref{fig:delta} (see Supplemental Material\,\cite{supp}, Fig. S8), is shown in Fig.\,\ref{fig:chi}, alongside susceptibility data from elastoresistivity measurements reported in Ref.\,\cite{eckberg_sixfold_2020}.

The susceptibility curves obtained from x-ray and elastoresistivity studies show remarkable agreement (Fig.\,\ref{fig:chi}). Both datasets exhibit a similar profile: a nonzero response emerges around $\sim T_\mathrm{IC}$,  increases with decreasing temperature in the I-CDW regime following a Curie-Weiss trend, and vanishes at the triclinic transition. Additionally, our values of $\chi(T)$ for $x = 0.27$ fall almost exactly between those measured for $x = 0.2$ and $x = 0.4$ in elastoresistivity studies (Fig.\,\ref{fig:chi}). The strong correlation between x-ray and transport susceptibilities provides compelling evidence for a direct correspondence between the nematic order associated with I-CDW anisotropy and that observed in transport measurements. This observation also suggests that the nematic order originates from the I-CDW degrees of freedom.

We emphasize that the preservation of tetragonal symmetry in the I-CDW phase\,\cite{lee_multiple_2021} is crucial for defining a nematic susceptibility from x-ray measurements. The translational symmetry-breaking potential imposed by the 2Q I-CDW order on the electronic degrees of freedom likely induces changes in the electronic spectrum, resulting in the nematic instability observed in transport studies\,\cite{eckberg_sixfold_2020}. This interpretation is supported by the fact that $\chi(T)$ remains essentially zero above $T_\mathrm{IC}$. 

Analyzing the symmetries within a simple Landau free energy framework provides insight into the coupling between the CDW and strain. To this end, we consider the Landau free energy for the I-CDW order parameter\,\cite{nie_vestigial_2017,fernandes_intertwined_2019}, 
\begin{align}
    F = F_0 + F_1
    \label{eq:free_en}
\end{align}
where
\begin{equation}
    F_0 = \frac{a}{2}(|\rho_x|^2 + |\rho_y|^2) + \frac{b}{4}(|\rho_x|^2 + |\rho_y|^2)^2 + \frac{c}{2}(|\rho_x|^2 - |\rho_y|^2)^2,
\end{equation}
is the I-CDW free energy and
\begin{equation}
    F_1 = - \lambda_\rho(|\rho_x|^2 - |\rho_y|^2)\epsilon,
\end{equation}
describes the coupling between the I-CDW and strain. Here, $\rho_x$ and $\rho_y$ are complex amplitudes that characterize the I-CDW modulation along the $x$- and $y$- directions, i.e., the in-plane charge density is given by $\rho=\rho_0 + \rho_x e^{i\mathbf{Q}\cdot\mathbf{\hat{x}}} + \rho_y e^{i\mathbf{Q}\cdot\mathbf{\hat{y}}} + C.C$ with 
$|\rho_x|^2\propto I_x$ and $|\rho_y|^2\propto I_y$.  The parameter $a(T) = a_0(T – T_\mathrm{IC})$ denotes the proximity to the I-CDW transition temperature. The B$_{1g}$ strain is given by $\epsilon_{xx}- \epsilon_{yy} = \epsilon_{xx} (1 + \nu)=\epsilon$, where $\nu$ represents the Poisson ratio for the piezo stack. We approximate $\nu$ as constant over the temperature range investigated.

The CDW-strain coupling function, $\lambda_\rho$, can be expanded in powers of the (symmetric) CDW order parameter $|\rho|^2 \equiv |\rho_x|^2 + |\rho_y|^2 $:
\begin{equation}
    \lambda_\rho = \lambda + \lambda' |\rho|^2 + \mathcal{O}(|\rho|^4).
\end{equation}
In most approaches, only the leading-order (harmonic) term is retained\,\cite{fernandes_what_2014}. The primary effect of this term, $\lambda$, is to split the transition temperatures of the $x$- and $y$-modulated CDWs. However, the data in Fig.\,\ref{fig:icdw_t} show no such splitting, suggesting the need to include the higher-order coupling term, $\lambda'$. This term introduces anharmonic coupling, as the lattice distortion $\epsilon$ induced by unequal intensities at the two I-CDW wave vectors scales with the intensity squared, $|\rho|^4$, rather than linearly with $|\rho|^2$. A related anharmonic term has been proposed to explain the nature of the coupled nematic and magnetic transitions in the related iron-based compound BaFe$_2$As$_2$\,\cite{kim_character_2011}.

Incorporating this additional coupling term, our total free energy (Eq.\,\ref{eq:free_en}) successfully captures all features of the data for $T>135$\,K, above the triclinic transition. We overlay these free energy fits as solid lines in Fig.\,\ref{fig:icdw_t}. First, we find that $c>0$, consistent with the preservation of tetragonal symmetry in the I-CDW phase. Second, we determine that $\lambda$ and $\lambda'$ have opposite signs and similar magnitudes, with $\lambda=39(3)$ and $\lambda'=-82(1)$. This result shows that anharmonic effects are essential to fully describe the influence of strain on the CDW degrees of freedom in BSNA. 

Beyond establishing the role of anharmonic CDW-strain effects in BSNA, our analysis provides insights into the mechanism underlying the anisotropic resistivity associated with nematicity. Previously, the strong temperature dependence of the elastoresistance inside the I-CDW phase was attributed to an electronic contribution to the nematic order parameter\,\cite{eckberg_sixfold_2020}. Here, we showed that although the I-CDW phase does not break tetragonal symmetry, it hosts a sizable nematic susceptibility. This is also the case of the double-Q spin-density wave observed in the doped iron-pnictide compound CaKFe$_4$As$_4$\,\cite{bohmer_evolution_2020}. Moreover, similar to the spin degree of freedom in their iron-based counterparts, the charge degree of freedom in BSNA couples to strain through a bilinear term involving the antisymmetric component of charge density fluctuations, $(|\rho_x|^2-|\rho_y|^2)$. A key difference, however, is that the nematicity in BSNA also depends on the symmetric combination, $(|\rho|^2)$. This additional dependence means that, unlike the iron-based spin-driven nematic systems, the overall and relative magnitudes of the density fluctuations contribute significantly to the nematicity of Ba$_{1-x}$Sr$_x$Ni$_2$As$_2$.

Since the nematic susceptibility only appears in the I-CDW regime, the resistivity anisotropy likely arises from the strain-induced difference in the two CDW $\rho_x$ and $\rho_y$ order parameters, implying a charge-driven nematicity. We postulate that this difference drives an anisotropic reconstruction of the Fermi surface, causing inequivalent averaged Fermi velocities along the $x$ and $y$ directions. The resulting Drude weight anisotropy manifests as an anisotropy in the DC resistivity. The similarity between the temperature dependence of the I-CDW nematic susceptibility $\partial \Delta/\partial \epsilon$ and elasto-resistivity measurements provides further evidence for this scenario.

Finally, we note that nematic fluctuations in BSNA have been associated with the sharp enhancement of the superconducting transition temperature observed upon substitution\,\cite{eckberg_sixfold_2020}. Our results demonstrate a significant anharmonic coupling between charge fluctuations and electronic nematicity and therefore raise the intriguing question of whether this anharmonic coupling contributes to the enhancement of $T_c$.


\textit{Acknowledgments}---This work was primarily supported by the Center for Quantum Sensing and Quantum Materials, an Energy Frontier Research Center funded by the U.S. Department of Energy (DOE), Office of Science, Basic Energy Sciences (BES), under Award No. DE-SC0021238. Growth of BSNA crystals was supported by the AFOSR Grant No. FA9550-14-10332 and the National Science Foundation Grants No. DMR1905891 and No. DMR2225920, as well as the Maryland Quantum Materials Center and the Maryland Nanocenter and its FabLab. P.A. and J.P.P. gratefully acknowledge additional support from the EPiQS program of the Gordon and Betty Moore Foundation, Grants No. GBMF9452 and GBMF9071, respectively.

\nocite{*}

\bibliography{BSNA}

\end{document}